%% file: main.tex
\documentclass[conference]{IEEEtran}
\usepackage{graphicx,color} 
\usepackage{xcolor,url,xurl}
\usepackage{dirtytalk}
\usepackage{hyperref}
\usepackage{wrapfig}
\usepackage[nocompress]{cite}
\newcommand{\ignore}[1]{}

\newtheorem{insight}{{\bf Observation}}

\title{Space Cybersecurity Testbed:  Fidelity Framework, Example Implementation, and Characterization}

\author{\IEEEauthorblockN{Jose Luis Castanon Remy,
Caleb Chang,
Ekzhin Ear,
and 
Shouhuai Xu
}
\IEEEauthorblockA{\{jcastano, cchang, eear, sxu\}@uccs.edu\\
Laboratory for Cybersecurity Dynamics, Department of Computer Science, University of Colorado Colorado Springs}}
\date{}
\begin{document}
\bstctlcite{IEEEexample:BSTcontrol}


\IEEEoverridecommandlockouts
\makeatletter\def\@IEEEpubidpullup{6.5\baselineskip}\makeatother
\IEEEpubid{\parbox{\columnwidth}{
		{\fontsize{7.5}{7.5}\selectfont Workshop on Security of Space and Satellite Systems (SpaceSec) 2025 \\
			24 February 2025, San Diego, CA, USA \\
			ISBN 979-8-9919276-1-1 \\
			https://dx.doi.org/10.14722/spacesec.2025.23042  \\
			www.ndss-symposium.org}
}
\hspace{\columnsep}\makebox[\columnwidth]{}}

\maketitle

\begin{abstract}
Cyber threats against space infrastructures, including satellites and systems on the ground, have not been adequately understood. Testbeds are important to deepen our understanding and validate space cybersecurity studies. The state of the art is that there are very few studies on building testbeds, and there are few characterizations of testbeds. In this paper, we propose a framework for characterizing the {\em fidelity} of space cybersecurity testbeds. The framework includes 7
attributes for characterizing the system models, threat models, and defenses that can be accommodated by a testbed. We use the framework to guide us in building and characterizing a concrete testbed we have implemented, which includes space, ground, user, and link segments. In particular, we show how the testbed can accommodate some space cyber attack scenarios that have occurred in the real world, and discuss future research directions.

\end{abstract}


\section{Introduction}

Space cybersecurity is an important, yet underdeveloped, technical field.
In particular, the cybersecurity community needs a better understanding of space systems because space cybersecurity is not adequately understood.
This is not surprising because space systems are highly technical and out of the scope of the traditional cybersecurity or computer science curriculum. One particular void is the lack of realistic experimental testbeds whereby we can conduct space cyber attack and defense research to observe phenomena and collect data to help build advanced models. 
That is, we need high-quality testbeds to support and validate space cybersecurity research. The state of the art is that there are few proposals on building such testbeds (e.g., \cite{finke2023satellite}). Moreover, there is no systematic understanding of {\em what} constitutes high-quality testbeds. For instance, there is no systematic way to describe space cybersecurity testbeds, such as the threat models they can support. This motivates the present study.

\smallskip

\noindent{\bf Our Contributions}. This paper makes two contributions. First, we propose the first framework for describing the {\em fidelity} of space testbeds. The framework defines 7 attributes 
to characterize the system models, threat models, and defenses that can be accommodated by a testbed. These attributes offer a systematic way to describe space cybersecurity testbeds, including a segment-component-module-element representation of space testbed hardware fidelity and a method for using {\em functions} to describe the implementations of space {\em missions} to enable space cyber risk analysis.  Second, we show that the framework can guide the design and implementation of testbeds, by presenting a concrete a 4-segment testbed guided by the framework. We characterize the testbed via the 7 attributes. Our observations include: (i) open-source flight software is prone to cyber attacks; (ii) testbeds can help identify elements for hardening purposes; and (iii) mission-specific cyber risk analysis can guide mission risk management. 

\smallskip

\begin{figure*}[!htbp]
\centering{\includegraphics[width=1.5\columnwidth]{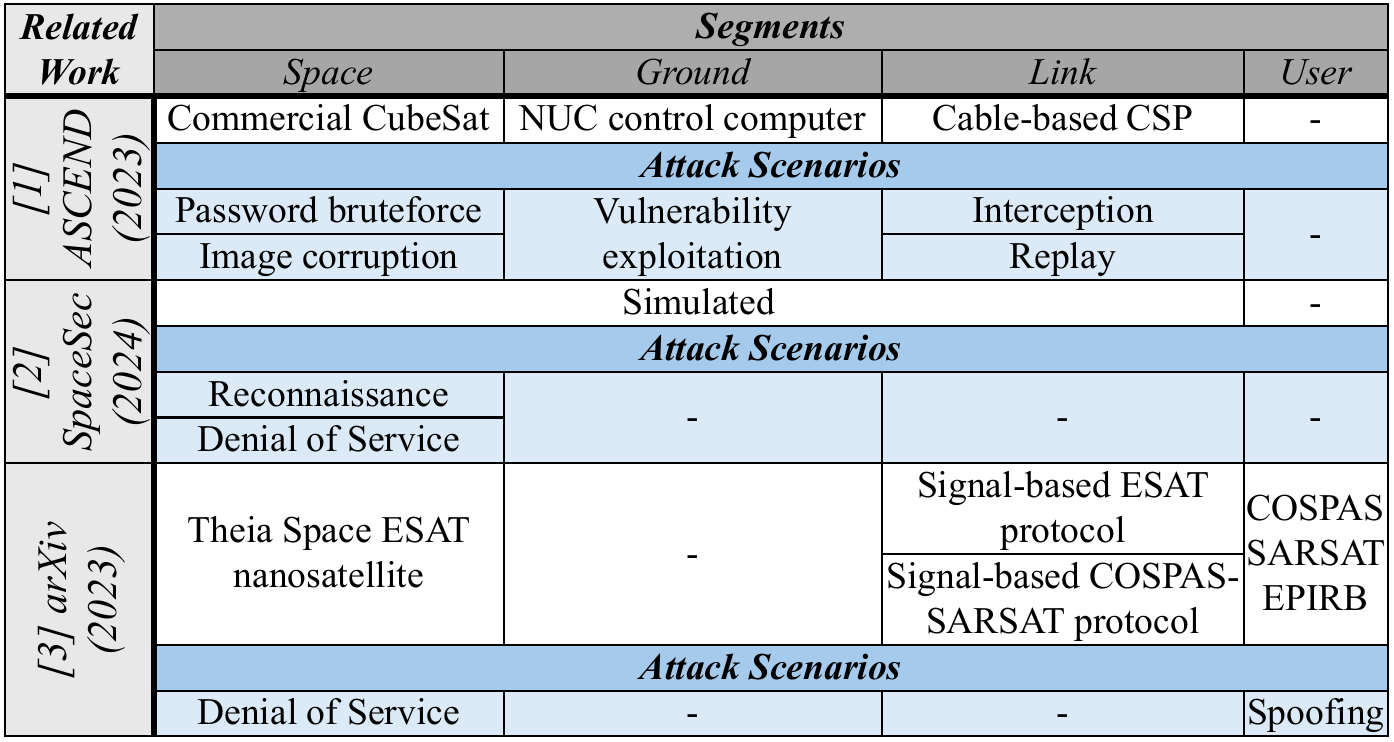}}
\caption{Related work comparison. NUC = Next Unit Computer; CSP = CubeSat Space Protocol; COSPAS-SARSAT = Space System for the Search of Vessels in Distress Search And Rescue Satellite-Aided Tracking; EPIRB = Emergency Position Indicating Radio Beacon; ESAT = Educational Satellite.}
\label{fig:relwtable}
\end{figure*}

\noindent{\bf Related Work}.
To our knowledge, there are only 3 space testbeds reported in the literature \cite{finke2023satellite, collinsmerge, costin2023towards}. Figure \ref{fig:relwtable} summarizes the three testbeds, including the attack scenarios they can accommodate at specific segments. 
When compared with these three testbeds, ours has the following advantages: (i) our satellite (i.e., CubeSat) has multiple payloads rather than a single payload; (ii) our satellite is highly customizable, for instance, allowing us to modify flight software and add additional payloads; (iii) our testbed has a realistic 
link segment that employs Radio Frequency (RF) communications rather than using USB cables; (iv) our testbed does not use any simulated segment but rather uses real systems and channels; 
(v) our testbed has a user segment, which is only present in one of the three existing testbeds; and 
(vi) our testbed has a ground segment with functions for controlling satellites, receiving telemetry from satellites, and controlling and receiving data from the camera payload.

\smallskip

\noindent{\bf Paper Outline}. 
Section \ref{sec:framework} presents the framework for describing space cybersecurity testbed fidelity.
Section \ref{sec:testbed}
describes and characterizes a concrete testbed that is built under the guidance of the framework.
Section \ref{sec:discussion}
discusses future research directions. Section \ref{sec:conclusion}
concludes the paper.

\section{Fidelity Framework}
\label{sec:framework}

To guide the design and characterization of space cybersecurity testbeds, we propose a framework for describing the {\em fidelity} of their 
system models, threat models, and defenses. 
These three aspects are essential to investigate space cybersecurity.
We define three kinds of attributes: (i) attributes that characterize system models, showing how comprehensive a space testbed is; (ii) attributes that characterize threat models, showing the kinds of attacks that can be accommodated by a testbed;
and (iii) attributes that characterize defenses, showing the kinds of defenses that can be accommodated by a testbed. 

\begin{figure}[!htbp]
\centering{\includegraphics[width=\columnwidth]{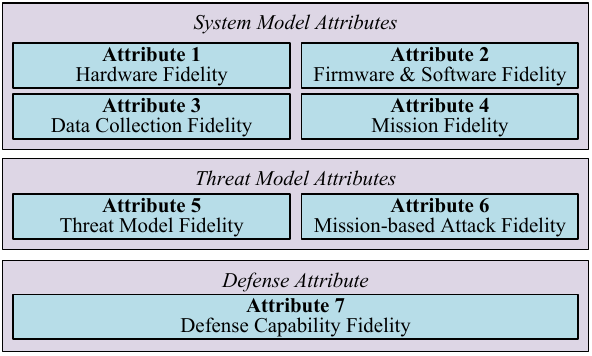}}
\caption{The fidelity framework with 7 attributes.
}
\label{fig:framework}
\end{figure}

Figure \ref{fig:framework} summarizes the 7 attributes defined in the framework, including 4 attributes for characterizing system models, 2 attributes for threat models, and 1 attribute for defenses. These attributes are elaborated below.


\subsection{System Model Attributes}

\noindent{\bf Attribute 1: Hardware Fidelity}.
A {\em space infrastructure} typically consists of 4  segments, known as space, link, ground, and user. Note the launch segment, sometimes called the fifth segment, is not considered in this study. Other than the link segment, any segment can be divided into multiple {\em components}; e.g., a space segment often consists of a {\em bus system} component and a {\em payload} component.
A component can be decomposed into multiple {\em modules}. For instance, the bus system component often consists of the following modules: {\em propulsion}, {\em communications}, {\em on-board computer (OBC)}, {\em attitude and orbit determination (AOD)}, and {\em electrical power}. A module can be further decomposed into multiple {\em elements}; e.g., the communications module often consists of two elements: {\em software-defined radio} and {\em antenna}.
Element is the finest granularity in the present paper, where every element consists of a hardware, a firmware, or a software implementation. 
Note that segments, components, modules, and elements formulate a hierarchical structure.
This segment-component-module-element representation offers a systematic description of space infrastructure hardware.
Correspondingly, the {\em hardware fidelity} attribute describes the segments, components, modules, and elements 
that are accommodated by a space cybersecurity testbed.

\smallskip

\noindent{\bf Attribute 2: Firmware and Software Fidelity}. 
A space infrastructure also consists of firmware and software, which typically run in some elements. Note that elements not running any firmware or software (e.g., sensor) are considered pure hardware elements, which are described by Attribute 1. 
This attribute, {\em firmware and software fidelity}, describes the firmware and software elements that are accommodated by a testbed.

\smallskip

\noindent{\bf Attribute 3: Data Collection Fidelity}. 
One important purpose of building space cybersecurity testbeds is to collect data. Data generated by a testbed includes: 
hardware data, such as sensor data (e.g., satellite temperature); firmware data, such as satellite microcontroller logs; software data, such as flight software logs; and mission data, such as images or text communications.
From a different perspective, we can divide data into two categories: space-related data, such as telemetry data (e.g., satellite temperature) and payload data (e.g., images); and cyber-related data, such as mission software logs. 
This attribute, {\em data collection fidelity}, describes the kinds of data that can be generated by, and collected from, a space cybersecurity testbed from the preceding two perspectives. 

\smallskip

\noindent{\bf Attribute 4: Mission Fidelity}. 
We specify space missions in two kinds: (i) application-level missions that are supported or enabled by a space infrastructure, where different missions may demand different kinds of support; 
(ii) infrastructure-level missions that support application-level missions.
We describe missions via functions, namely how missions are implemented at the aforementioned element-level of abstraction. This is important because one mission may be implemented by a single function or jointly by multiple functions.
Functions, which pave a way for conducting mission-specific cyber risk analysis, 
are represented as a graph-theoretic structure (i.e., {\em directed graph}), where nodes represent elements and arcs represent how information (i.e., command or data) is transmitted between nodes.
Correspondingly, there are two families of functions: the ones that implement application-level missions (e.g., using the payload camera to take images of the Earth); and
the ones that implement infrastructure-level missions (e.g., controlling a satellite). 
This attribute, {\em mission fidelity}, describes application-level and infrastructure-level missions (i.e., their functions) that can be accommodated by a testbed.

\subsection{Threat Model Attributes}

A testbed should be able to accommodate both the attacks that have occurred in the real world and the attacks that have not occurred but could occur in the future (e.g., new attacks described in academic literature).

\smallskip

\noindent{\bf Attribute 5: Threat Model Fidelity}.
This attribute characterizes the
attacks that can be accommodated by a testbed. We specify threat models via: (i)
{\em attack point}, namely the element that an attacker compromises as an entry point from outside a space infrastructure;
(ii) {\em attack vector}, which describes the {\em attack techniques} \cite{ATTCK, SPARTA} that can be used by an attacker to gain access to the attack point;
(iii) {\em vulnerability}, which is exploited by an attack;
and (iv) {\em attack consequence}, which describes the damage incurred by an attack, in terms of the elements that can be compromised by the attack.
The {\em threat model fidelity} attribute describes the kinds of threat models that can be accommodated by a testbed.



\smallskip

\noindent{\bf Attribute 6: Mission-based Attack Fidelity.} 
The {\em mission-based attack fidelity} attribute characterizes how attacks may disrupt missions in terms of the functions implementing them.
This is important because understanding attacks from the point of view of missions paves a way for analyzing cyber risks to missions through their functions, helping identify countermeasures to harden missions. 



\subsection{Defense Attribute}

\noindent{\bf Attribute 7: Defense Capability Fidelity}. 
The {\em defense capability fidelity} attribute describes the defenses that can be accommodated by a testbed. Defenses can be deployed at elements, modules, or components, analogous to host-based defense in Information-Technology (IT) systems. Defenses can also be deployed 
at communication channels between elements, modules, components, or segments, analogous to 
network-based defense in IT networks. 
Defenses can be described in terms of countermeasures to mitigate attack consequences \cite{ear2024towards}. Countermeasures include {\em security controls} designed for traditional IT infrastructures (e.g., \cite{NIST800-53r5}) or for space infrastructures (e.g., \cite{NASA-STD-1006A}).

\section{Our Testbed and Its Characterization}
\label{sec:testbed}

We have designed and built a testbed under the guidance of the framework, especially using Attributes 1-4 in an iterative process. Specifically, our design started with initial hardware (Attribute 1) and missions (Attribute 4). Then, firmware and software (Attribute 2) and data collection (Attribute 3) were developed to enable the missions. If the missions cannot be fulfilled, then Attributes 1-3 should be reassessed to ensure Attribute 4. Alternatively, Attribute 4 can be assessed once a testbed is designed and developed, then possibly incorporating additional missions  (e.g., a remote sensing payload like a camera could be used to capture additional mission data by observing different spectrum physical phenomena). 
We have used the testbed to conduct successful cyber attack and defense experiments, illustrating Attributes 5-7. 
We describe our testbed via the aforementioned 7 attributes.

\subsection{System Model Attributes Characterization of Our Testbed}

\subsubsection{Attribute 1: Hardware Fidelity}
\begin{figure}[!htbp]
\centering{\includegraphics[width=.8\columnwidth]{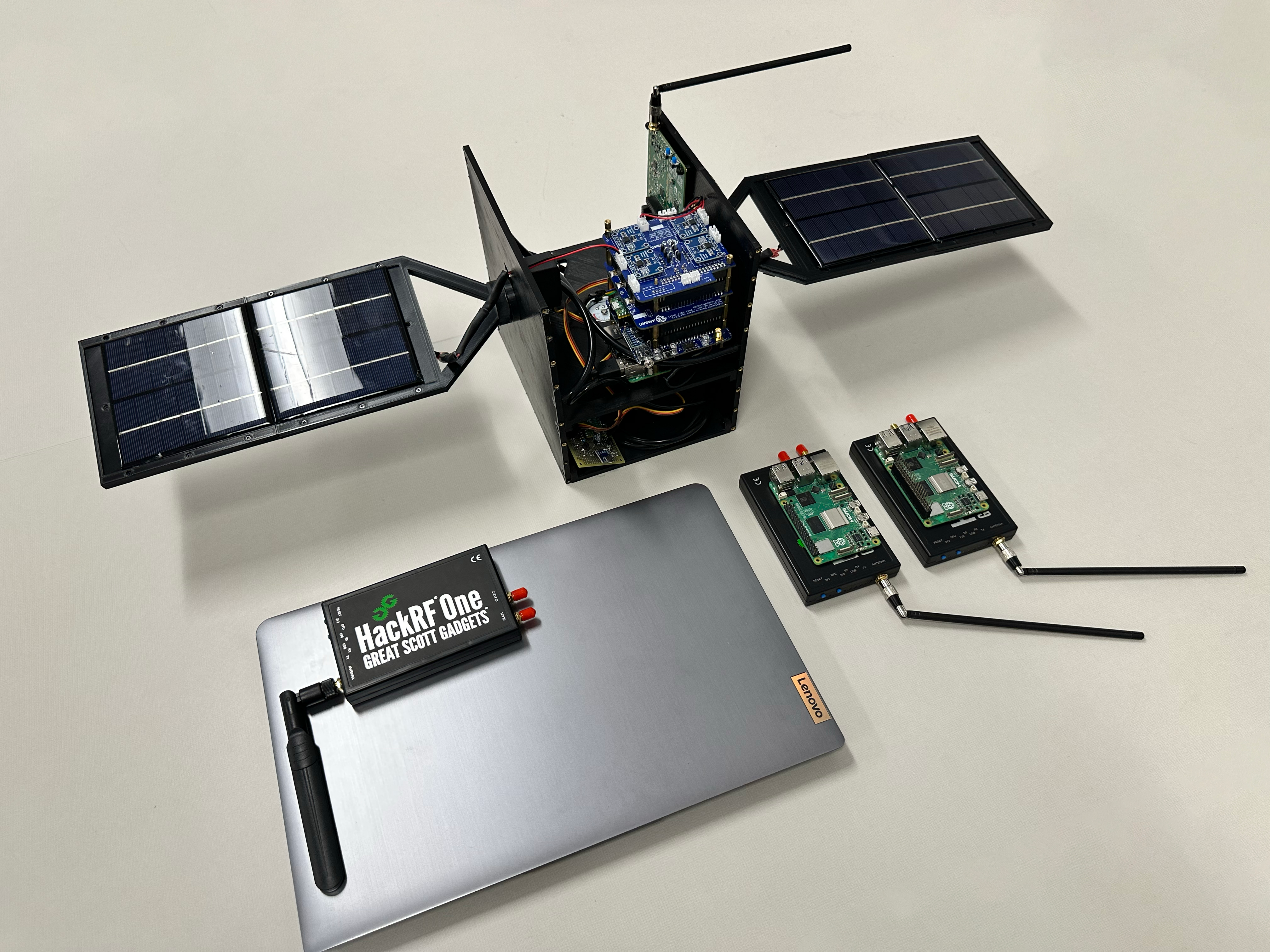}}
\caption{Our testbed consists of one CubeSat, one ground station (including software-defined radio, shown at bottom left), and two user terminals (with software-defined radio, shown on the right).} 
\label{fig:implementation}
\end{figure}

Figure \ref{fig:implementation} depicts our testbed, which consists of a space segment
with one CubeSat satellite, a ground segment with one ground station, a user segment with two user terminals, and a link segment with RF communication channels between the 
segments. The testbed currently supports: (i) one infrastructure-level mission, {\em bus management}, which controls and monitors the health status of the satellite; and (ii) two application-level missions, {\em remote sensing} that provides visual imagery collection
and {\em communications} 
whereby user terminals can send and receive audio, text, and binary messages between each other. 

\begin{figure*}[!htbp]
\centering{\includegraphics[width=\textwidth]{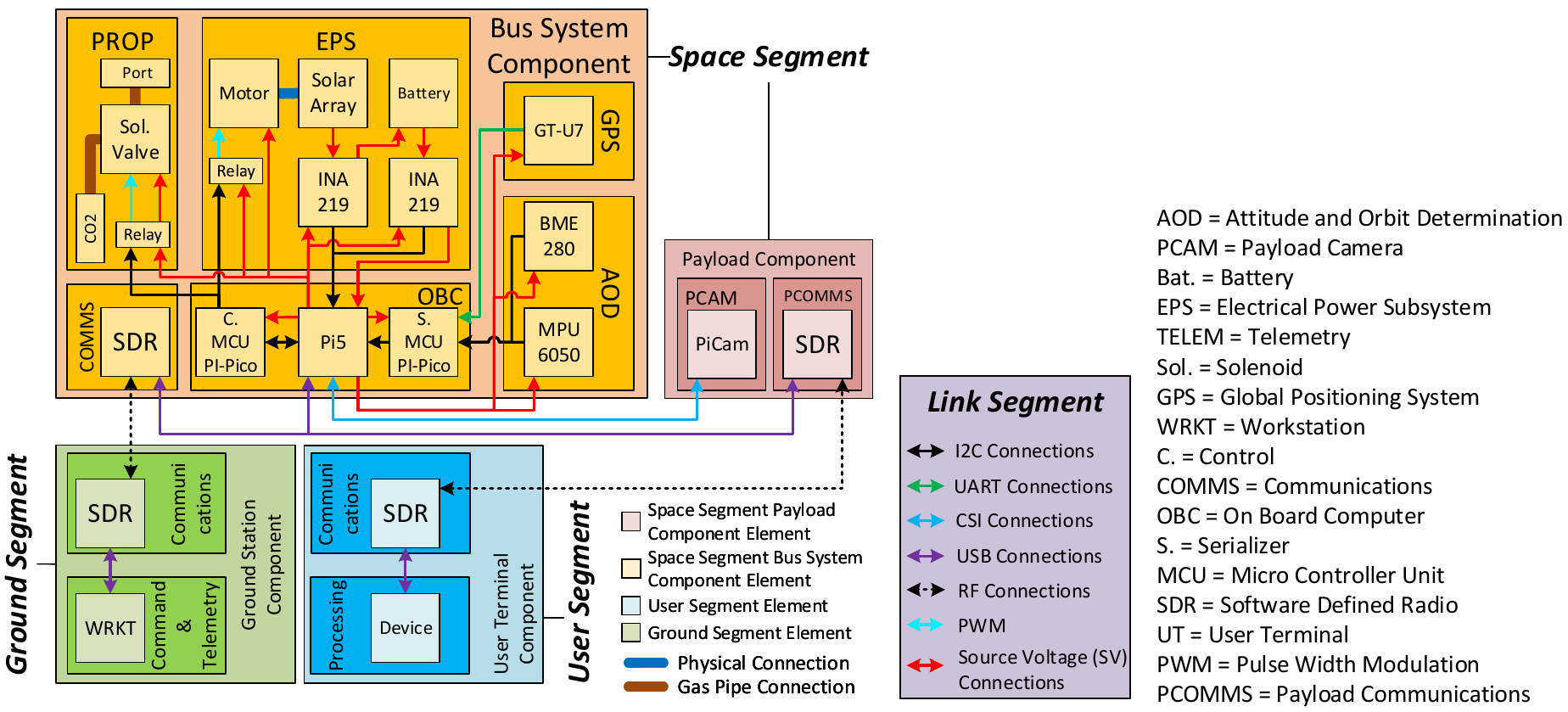}}
\caption{Hardware-fidelity
in the segment-component-module-element representation.
}
\label{fig:system_model_element_level}
\end{figure*}

Figure \ref{fig:system_model_element_level} highlights the segment-component-module-element representation of the testbed. We characterize the
hardware fidelity of the testbed with respect to the 4 segments.



\smallskip

\noindent{\bf Space Segment}. 
This segment has two components: payload and bus system. 
The payload enables the two missions mentioned above, namely remote sensing and communications.
The bus system 
controls the satellite's position, and monitors and 
maintains the health condition of the satellite. 

The payload component has two modules: {\em Payload Camera (PCAM)}, which takes images (i.e., pictures) via its {\em PiCam} element;
and {\em Payload Communications (PCOMMS)}, which receives RF signals from (via its {\em SDR} element), and transmits RF signals to, the {\em SDR} element in the {\em user} segment. 

The bus system component has 6 modules: {\em Communications (COMMS)}, {\em Propulsion (PROP)}, {\em Electrical Power Subsystem (EPS)}, {\em OBC}, {\em AOD}, and {\em Global Positioning System (GPS)}.
(i) The COMMS module
receives RF signals from, and transmits RF signals to, the {\em ground} segment via its {\em SDR} element.
(ii) The PROP module maneuvers the satellite 
(i.e., orbit modifications) and provides accurate positioning (i.e., attitude, pitch, and roll modifications) via its {\em solenoid valve} element, which controls the amount of CO2 propelled. 
(iii) The EPS module provides electrical energy to the other modules in the bus system and payload components via its {\em solar array} element, which produces electrical energy that is stored in the {\em battery} element. The {\em INA219} element provides voltage and current telemetry data of the solar array and battery elements. 
Note that the {\em motor} element controls the rotation of the solar array, optimally directing it
towards the sun for maximum energy generation.
(iv) The AOD module provides temperature, humidity, pressure, altitude, acceleration, and rotation speed telemetry data.
(v) The GPS module provides altitude, longitude, and latitude telemetry data.
(vi) The OBC module leverages the {\em Pi5} element to:
(a) control the modules within the bus system and payload components 
by interpreting commands received from the ground segment; 
and (b) maintain, serialize, and synchronize telemetry data generated in the PROP, EPS, AOD, and GPS modules. 
Note that the Pi5 element leverages the {\em Control Micro Controller Unit (C. MCU PI-Pico)} element to control the solenoid valve and motor elements, while using the {\em Serializer Micro Controller Unit (S. MCU PI-Pico)} element to serialize telemetry data generated in the AOD and GPS modules.

\smallskip

\noindent{\bf User Segment}.
As highlighted in Figure \ref{fig:system_model_element_level}, the user segment has a {\em user terminal} component that enables the communications mission mentioned above. The user terminal component has two modules: {\em communications} and {\em processing}. The communications module receives RF signals from, and transmits RF signals to, the space segment via the {\em software-defined radio (SDR)} element of this module. The processing module provides a user interface for an end user to send and receive audio, text, and binary messages via its {\em device} element. 

\smallskip

\noindent{\bf Ground Segment}.
As highlighted in Figure \ref{fig:system_model_element_level}, the ground segment has a {\em ground station} component 
that (i) enables the remote sensing mission mentioned above
by commanding the {\em PCAM} module 
to capture images, 
and (ii) controls the {\em bus system} component of the satellite. 
The ground station component has two modules: {\em communications} and {\em command and telemetry}.
The communications module receives RF signals from, and transmits RF signals to, the space segment via its {\em SDR} element. 
The command and telemetry module leverages its {\em workstation (WRKT)} element to achieve the following: (i) command and control the payload of the satellite by instructing the PCAM to take a picture;
(ii) command and control the bus system of the satellite by instructing the motor to rotate the solar array;
(iii) command and control the bus system of the satellite by instructing the PROP to change the orbit of the satellite;
(iv) analyze the images captured 
by the PCAM 
of the satellite; and
(v) monitor the health of the satellite by analyzing telemetry data received from the bus system component.

\smallskip

\noindent{\bf Link Segment}.
As highlighted in Figure \ref{fig:system_model_element_level}, the link segment connects the elements of the same or different segments. Communications between elements in the same segment are cable-based connections with different types of data transfers (e.g., the WRKT and the SDR elements are connected through a USB cable). Communications between elements in different segments are referred to as {\em downlink} (i.e., space segment to ground or user segment) and {\em uplink} (i.e., ground or user segment to space segment). Downlink and uplink communications use RF transmissions on Multi-Use Radio Service (MURS) VHF channels. Specifically, space-to-ground downlinks and ground-to-space uplinks use 151.820 MHz with 11.25kHz of bandwidth; space-to-user downlinks use 154.570 MHz with 20kHz of bandwidth; and user-to-space uplinks use 154.600 MHz also with 20kHz of bandwidth. Note that SDR transmission power is set minimally to practice responsible RF emissions, and SDRs are connected with coaxial cables to contain RF emissions when licensed frequencies are used.

\subsubsection{Attribute 2: Firmware and Software Fidelity} This attribute of the testbed is characterized as follows.

\smallskip

\noindent{\bf Firmware Fidelity}. The following elements can run firmware. 
(i) The C. MCU Pi-Pico element in the OBC module of the bus system component of the satellite runs a firmware that interprets and executes propulsion and rotation commands received from the Pi5 element in the OBC module. 
(ii) The S. MCU Pi-Pico element in the OBC module 
runs a firmware that receives telemetry data from the {\em GT-U7}, {\em BME280} and {\em MPU6050} elements and serializes the received data.
(iii) The {\em PiCam} element in the PCAM module of the payload component of the satellite runs a firmware that interprets and executes the commands received from the Pi5 element. 
(iv) The GT-U7 element in the GPS module of the bus system component of the satellite runs a firmware that converts the received GPS RF signals into location data.
(v) The SDR elements run a firmware to 
demodulate the received RF signals.
Note that the C. MCU Pi-Pico and S. MCU Pi-Pico elements run tailored firmware, while the rest of the modules run firmware provided by commercial vendors.

\smallskip

\noindent{\bf Software Fidelity}. The following elements can run software.
(i) The Pi5 element 
runs the open-source core Flight System (cFS) software \cite{cfsdocumentation}, which is a layered flight software that 
controls and manages the elements of the bus system component.
The Pi5 also runs the open-source GNU Radio software \cite{GNURadiodocumentation}, which processes RF signals 
through the SDR element in the bus system.
Note that both cFS and GNU Radio run on top of the operating system, 
which is supplied by the 
vendor.
(ii) The WRKT element in the command and telemetry module of the ground station component of the ground segment runs the open-source Comprehensive Open-architecture Solution for Mission Operations Systems (COSMOS) software \cite{cosmosdocumentation}, which controls the ground station component by producing commands and displays telemetry data received from the satellite. 
The WRKT element also runs 
GNU Radio which 
interacts with COSMOS.
(iii) The device element in the processing module of the user terminal component of the user segment runs tailored software to permit an end user to enter and visualize messages. The device element also runs 
GNU Radio which 
interacts with the tailored software running in the same module.

During the process of building our testbed, we observed the flight software, cFS, has many significant errors, such as segmentation faults that could facilitate memory injection attacks. This is likely due to: (i) the large number of software dependencies and third-party libraries that make cFS complex; (ii) the bespoke requirements for the flight software to operate with special-purpose hardware and other systems that prevents standardized testing of cFS; and (iii) developers' intent to make cFS ubiquitous for all flight missions by using cFS as a base for commercial flight software.
In summary, we make the following:
\begin{insight}
\label{insight:buggy software}
The open-source cFS
is prone to cyber attacks.
\end{insight}


\subsubsection{Attribute 3: Data Collection Fidelity}
The testbed generates hardware and mission data for various purposes.

\smallskip

\noindent{\bf Collecting Hardware-Generated Data}. 
The following 7 elements can collect hardware data.
(i) The INA219 element in the EPS module of the bus system component 
can collect current and voltage telemetry data generated by the solar array. 
(ii) The INA219 element in the EPS module of the bus system 
can collect current and voltage telemetry data generated by the battery. 
(iii) The Pi5 element in the OBC module of the bus system 
can collect data related to: the motor rotation position; the volume of the propulsion gas left in the tank;
the flight software and operating system logs;
and 
the telemetry data received by the other elements of the bus system component. 
(iv) The GT-U7 element in the GPS module of the bus system 
can collect GPS location telemetry data. 
(v) The BME280 element in the AOD module of the bus system 
can collect temperature, humidity, pressure and altitude telemetry data. 
(vi) The MPU6050 element in the AOD module of the bus system 
can collect acceleration and rotation speed telemetry data.
(vii) The S. MCU Pi-Pico element in the OBC module of the bus system 
processes, and thus can collect, telemetry data received from the GT-U7, BME280, and MPU6050 elements.

\smallskip

\noindent{\bf Collecting Mission-Generated Data}.
The following elements can collect mission-generated data.
(i) The PiCam element in the PCAM module of the payload component of the satellite can collect images. 
(ii) The WRKT element in the command and telemetry module of the ground station component of the ground segment processes and stores, and thus can collect, images. 
(iii) The device element in the processing module of the user terminal component of the user segment collects, processes, stores, and thus can collect, end user communication messages.
(iv) The Pi5 element 
processes, and thus can collect, end user messages 
and images. 




\subsubsection{Attribute 4: Mission Fidelity Characterization}

The testbed accommodates one infrastructure-level mission, known as bus management, and two application-level missions, namely remote sensing and communications. 
As described in the framework, missions are described by the functions that implement them. The bus management mission is implemented by two functions, {\em satellite control} (Function 1 below) and {\em telemetry downlink} (Function 2 below). Remote sensing is implemented by {\em Image Capture} (Function 3) and communications by {\em Message Transfer} (Function 4). 

\begin{figure}[!htbp]
\centering{\includegraphics[width=\columnwidth]{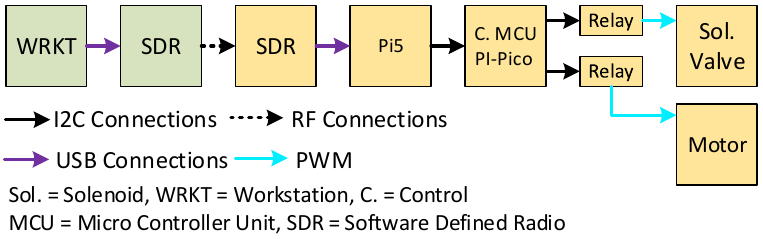}}
\caption{Satellite Control Function.}
\label{fig:bus_mngmnt_f}
\end{figure}

\noindent{\bf Function 1: Satellite Control}. 
As highlighted in Figure \ref{fig:bus_mngmnt_f}, this function describes how the ground station component controls the bus system component of the satellite. The function can be understood as follows.
(i) This function originates at the WRKT element in the command and telemetry module of the ground station component of the ground segment, by producing a propulsion or rotation command in the form of a digital signal.
(ii) The SDR element in the communications module of the ground station component of the ground segment modulates the digital signal into an RF signal and transmits the RF signal to the SDR element in the communications module of the satellite.
(iii) The SDR element demodulates the RF signal into a digital signal.
(iv) The Pi5 element in the OBC module of the bus system component of the space segment converts the digital signal into a command. 
(v) The C. MCU PI-Pico element in the OBC module of the bus system component of the satellite interprets the command and produces an electrical signal, which is then communicated to the intended element's relay. In the case of a propulsion command, the C. MCU PI-Pico element sends the electrical signal to the relay that is connected to the solenoid valve element, and the relay transforms the electrical signal into a pulse-width modulation (PWM) signal; in the case of a rotation command, the relay connected to the motor element receives the electrical signal.
(vi) The solenoid valve element receives the PWM signal and releases the propulsion gas (i.e., CO2) accordingly, producing thrust.
(vii) The motor element receives the PWM signal and rotates accordingly, changing the angle of the solar array.

\begin{figure}[!htbp]
\centering{\includegraphics[width=\columnwidth]{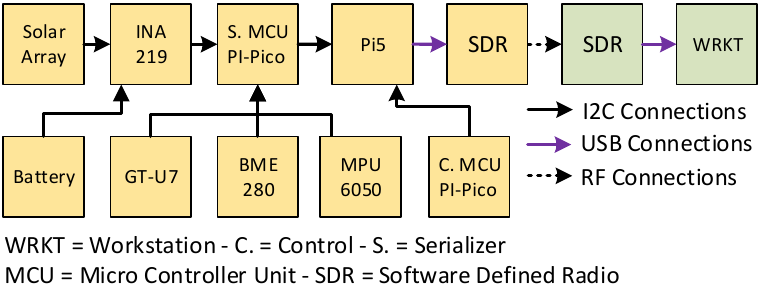}}
\caption{Telemetry Downlink Function. 
}
\label{fig:telem_downlink_f}
\end{figure}

\noindent{\bf Function 2: Telemetry Downlink}. 
As highlighted in Figure \ref{fig:telem_downlink_f}, this function describes how the ground station component receives telemetry data from the bus system component. It can be understood as follows.
(i) The function originates at the INA219 element in the EPS module, 
the C. MCU PI-Pico element in the OBC module, the GT-U7 element in the GPS module, the BME280 element in the AOD module, or the MPU6050 element in the AOD module; all these elements, and the modules to which they belong, are in the bus system component of the satellite. 
(ii) The S. MCU PI-Pico element in the OBC module receives telemetry data from the GT-U7, BME280, or MPU6050 element, and serializes the data.
(iii) The Pi5 element receives the serialized telemetry data from the S. MCU PI-Pico element. 
The Pi5 element serializes the received telemetry data with the motor rotation position and the volume of propulsion gas telemetry data, and then synchronizes and converts the received telemetry data into a digital signal.
(iv) The SDR element in the communications module (COMMS in Figure \ref{fig:system_model_element_level}) of the bus system component of the satellite modulates the digital signal into an RF signal and transmits it to the SDR element in the communications module of the ground segment.
(v) The SDR element demodulates the received RF signal into a digital signal.
(vi) The WRKT element in the command and telemetry module of the ground station component of the ground segment converts the digital signal into telemetry data, which is then analyzed to determine the health status of the satellite.

\smallskip

\begin{wrapfigure}{r}{0.36\textwidth}
\centering{\includegraphics[width=.72\columnwidth]{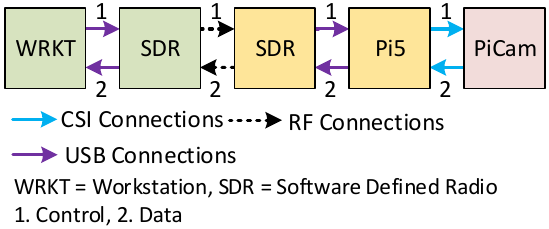}}
\caption{Image Capture Function.}
\label{fig:camera_control_f}
\end{wrapfigure}

\noindent{\bf Function 3: Image Capture}. This function fulfills the 
remote sensing mission, which provides surveillance of an area of the Earth determined by the orbit of the satellite and the field of view of the camera. This mission is important to understand space cybersecurity because an attacker could disrupt the mission by modifying the orbit of the satellite.
As highlighted in Figure \ref{fig:camera_control_f}, this function describes 
how the ground station component controls, and receives data from, the camera on the satellite. The function can be understood as follows.
(i) It originates at the WRKT element in the command and telemetry module of the ground station component of the ground segment, which produces an image-taking command in the form of a digital signal.
(ii) The SDR element in the communications module of the ground station component modulates the digital signal into an RF signal and transmits it to the SDR element in the communications module of the bus system component of the satellite. 
(iii) The SDR element demodulates the RF signal into a digital signal.
(iv) The Pi5 element in the OBC module 
converts the digital signal into a command.
(v) The Pi5 element commands the PiCam element to capture an image. 
(vi) The Pi5 receives the image 
from the PiCam element and converts them into a digital signal.
(vii) The SDR element in the communications module of the bus system component of the satellite modulates the digital signal into an RF signal and transmits the signal to the SDR element in the communications module of the ground station component.
(viii) The SDR element demodulates the RF signal into a digital signal.
(ix) The WRKT element converts the digital signal into an image. 


\smallskip

\noindent{\bf Function 4: Message Transfer}.
This function fulfills the
communications mission, which provides end users with text
communications capabilities
via handheld devices or user terminals in the same satellite footprint. This mission is relevant because an attacker could disrupt the mission 
by jamming or manipulating the communications.
\begin{figure}[!htbp]
\centering{\includegraphics[width=\columnwidth]{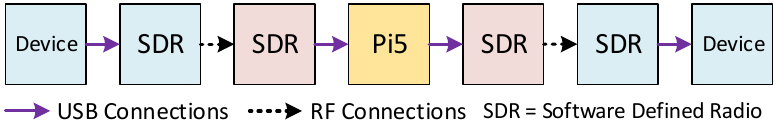}}
\caption{Message Transfer Function. 
}
\label{fig:comms_payload_f}
\end{figure}
As highlighted in Figure \ref{fig:comms_payload_f}, this function describes how one user terminal 
can communicate with another.
The function can be understood as follows.
(i) This function originates at the device element in the processing module of a user terminal,
where the end user can enter a message for another user and the message is converted into a digital signal.
(ii) The SDR element in the communications module of the user terminal
modulates the digital signal into an RF signal and transmits it to the SDR element in the communications module of the payload component of the satellite. 
(iii) The SDR element in the PCOMMS module of the payload component of the satellite demodulates the RF signal into a digital signal.
(iv) The Pi5 element in the OBC module 
processes the digital signal to be transmitted through another frequency as users may be assigned different frequencies to communicate with the satellite.
(v) The Pi5 sends the digital signal back to the SDR element in the payload communications module of the payload component of the satellite.
(vii) The SDR element modulates the digital signal into an RF signal and transmits it to another SDR element in the communications module of the user terminal.
(viii) The SDR element demodulates the RF signal into a digital one.
(ix) The device element converts the digital signal into a message, which is presented to the end user.

\begin{insight}
The Pi5 and SDR elements are essential to our testbed as they support all four functions.
\end{insight}

\subsection{Threat Model Attributes Characterization of Our Testbed}

We leverage the following 
real-world cyber attack, learned from \cite{fritz2013satellite, us2019report,ear2023characterizing}, to characterize our testbed in terms of Attributes 5 and 6. In 2008, an attacker interfered with NASA's Terra satellite RF communications twice; 2 minutes on June 20, 2008, and 9 minutes on October 22, 2008. 
Forensic analysis concluded that the attacker seized control of the Terra satellite bus system 
through an Internet connection at the SvalSat ground station intended for sharing payload data with users \cite{terrasattoday,terrahackgs, terrareuters}.
The attack inspires us to consider two threats: Threat 1 concerns the seizure of control of the Terra satellite; Threat 2 concerns the jamming of Terra's RF downlink transmission.

\subsubsection{Attribute 5: Threat Model Fidelity}

In  Threat 1, the attacker: 
(i) gains control of the satellite (attack consequence) 
through the WRKT element (attack point); 
(ii) gains access to the 
WRKT element via phishing techniques (attack vector); 
(iii) 
discovers a buffer-overflow vulnerability in the flight software deployed on the Pi5 element by using data gathered from the WRKT element (vulnerability); 
(iv) exploits the vulnerability
to remotely execute commands on
the satellite at the operating system level (attack vector); and (v) uses control of the satellite to 
disrupt
satellite operations and gather data (attack consequence). 
We have (partly) demonstrated Threat 1 in our testbed. 

In Threat 2, the attacker: 
(i) interferes with the downlink RF signals (attack point) to prevent the ground segment from receiving data from the space segment (attack consequence);
(ii) has access to a rogue terminal that is capable of producing RF signals with the appropriate frequency, timing, power, angle of arrival, and polarity to disrupt communications between the space and ground segments (attack vector); and
(iii) 
wages jamming attacks against the legitimate RF signals (attack consequence).
We have successfully demonstrated Threat 2 in our testbed.

\subsubsection{Attribute 6: Mission-based Attack Fidelity}
To characterize this attribute, we show that the aforementioned 
Threat 1 has an impact on Functions 1-4, and Threat 2 has an impact on Functions 2-3.

\ignore{
\begin{insight}
Attacks can be decomposed according to the functions where the impact of one (sub-)attack can be leveraged by another. 
\end{insight}
}



\begin{figure}[!htbp]
\centering{\includegraphics[width=\columnwidth]{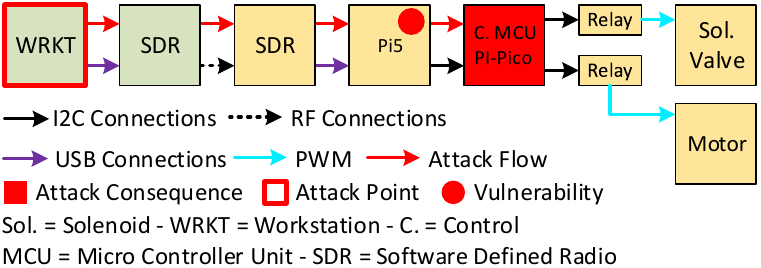}}
\caption{Threat 1 against Function 1 (satellite control)}
\label{fig:bus_mngmnt_af}
\end{figure}

\noindent{\bf Threat 1 against Function 1}. 
Figure \ref{fig:bus_mngmnt_af} highlights how Threat 1 can disrupt the satellite control function (Figure \ref{fig:bus_mngmnt_f}). Specifically, the attacker:
(i) enters the space infrastructure through the WRKT element of the ground segment;
(ii) uploads a malicious command packet containing exploit code to the WRKT element;
(iii) transmits the exploit code through the SDR element of the ground station component to the SDR element of the bus system component of the space segment, arriving at the Pi5 element;
(iv) executes the exploit code on the Pi5 element 
to bypass the flight software and gain direct access to the  C. MCU Pi-Pico element;
and (v) sends commands through the relay element to alter the states of the solenoid valve and motor elements to impact the physical operation of the satellite.




\begin{figure}[!htbp]
\centering{\includegraphics[width=\columnwidth]{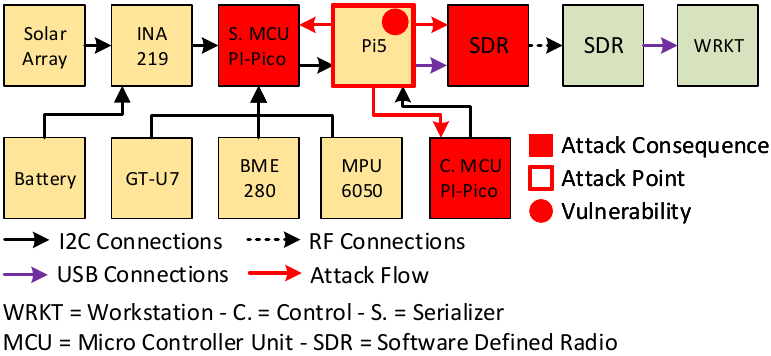}}
\caption{Threat 1 against Function 2 (telemetry downlink)}
\label{fig:telem_dwnlnk_af}
\end{figure}

\noindent{\bf Threat 1 against Function 2}.
Figure \ref{fig:telem_dwnlnk_af} highlights how Threat 1 can disrupt the telemetry downlink function (Figure \ref{fig:telem_downlink_f}). The attack begins at the Pi5 element, by gaining access to this element as described in Threat 1 against Function 1. 
Specifically, the attacker:
(i) commands the Pi5 element to manipulate the telemetry data received from the C. MCU Pi-Pico and S. MCU Pi-Pico elements; 
(ii) sends the manipulated data to the SDR element in the bus system component for transmitting it to the SDR element in the ground station;
and (iii) deceives the WRKT element with manipulated telemetry data, to corrupt the ground station's visibility of the satellite's operations (e.g., concealing the presence of the cyber attack).


\begin{figure}[!htbp] 
\centering{\includegraphics[width=.75\columnwidth]{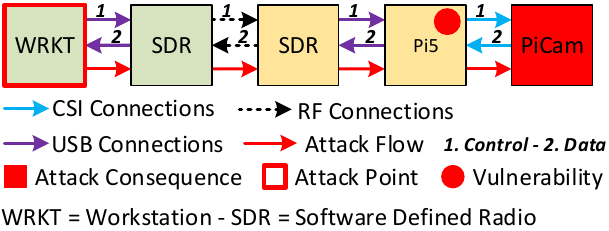}}
\caption{Threat 1 against Function 3 (image capture)}
\label{fig:camera_payload_af}
\end{figure}

\noindent{\bf Threat 1 against Function 3}.
Figure \ref{fig:camera_payload_af} highlights how Threat 1 can disrupt the image capture function (Figure \ref{fig:camera_control_f}). Specifically, the attack begins at
the WRKT element, where the attacker:
(i) crafts and transmits exploit code, similar to Figure \ref{fig:bus_mngmnt_af}, from the WRKT element to the Pi5 element to gain control of the Pi5;
and (ii) commands the PiCam element to adjust the shutter speed on the camera to an extreme value, rendering the PiCam element temporarily unable to capture images for the satellite's mission.





\begin{figure}[!htbp]
\centering{\includegraphics[width=.9\columnwidth]{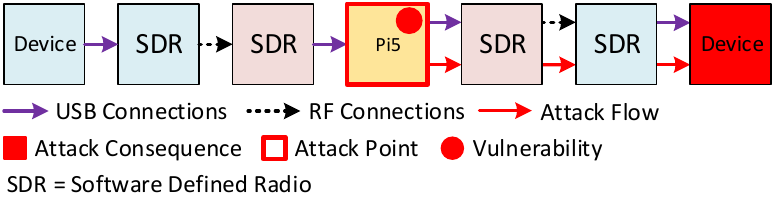}}
\caption{Threat 1 against Function 4 (message transfer)}
\label{fig:comms_payload_af}
\end{figure}

\noindent{\bf Threat 1 against Function 4}.
Figure \ref{fig:comms_payload_af} highlights how Threat 1 can disrupt the message transfer function (Figure \ref{fig:comms_payload_f}).
Specifically, the attack begins at the Pi5 element, where the attacker:
(i) intercepts packets sent from a device element in a user terminal component to the SDR element in the payload component of the satellite; 
(ii) manipulates the packets; 
and (iii) transmits the manipulated packets to the device element in another user terminal component, corrupting the communications between users.

\smallskip

\begin{wrapfigure}{r}{0.3\textwidth}
\centering{\includegraphics[width=.5\columnwidth]{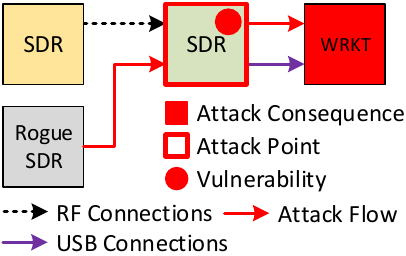}}
\caption{Threat 2 against Functions 2 and 3}
\label{fig:jamming_af}
\end{wrapfigure}

\noindent{\bf Threat 2 against Functions 2 and 3.} 
Figure \ref{fig:jamming_af} highlights how Threat 2 (jamming against downlink RF signals) can disrupt Functions 2 and 3, while noting that Functions 1 and 4 are not applicable because they do not involve the ground segment downlink.
We first describe what is common to both cases.
The attack proceeds as follows. (i) The attacker uses a 
{\em rogue SDR} element to transmit a malicious RF signal in the same frequency used between the SDR element of the satellite and the SDR element of the ground segment, which is referred to as {\em victim SDR}.
(ii) The victim SDR 
receives the malicious signal 
with a higher strength than the legitimate signal transmitted by the SDR element in the space segment. 
The malicious signal overlaps the legitimate signal, corrupting the transfer. The victim SDR receives the corrupt signal 
and demodulates it into a corrupt digital signal. Note that the malicious jamming signal could take the form of a cosine signal, which is a simple and cost-effective signal to produce for the attacker.
(iii) The WRKT element converts the corrupted signal into corrupted data, which takes the form of gibberish or randomly generated data.

The difference between the two cases is the following.
In the case of Function 2, the attacker prevents the victim from receiving telemetry data from the satellite, denying the ground station the ability to monitor the health status of the satellite and thus possibly destroying the satellite (e.g., if the satellite's temperature 
increases, the hardware could be damaged). 
In the case of Function 3, the attacker prevents the victim from receiving images collected by the satellite, which denies the remote sensing mission.








\subsection{Defense Attribute Characterization of Our Testbed}
Our testbed can employ traditional countermeasures in the ground segment (e.g., \cite{pavur2020sok,knez2016lessons, young2017commercial,vera2016cyber}), intrusion detection systems in the bus system component of the satellite (e.g., \cite{thebarge2022developing}), and jamming detection (e.g., \cite{borio2016jammer,querol2017real}) and RF interference mitigation (e.g., \cite{hu2023interference,lachapelle2021orbital,clements2023dual,falco2022space,turner2024jamming}) in the ground and space segments to defend the link segment.
We demonstrate in terms of Attribute 7 how our testbed can employ these countermeasures to address Threats 1 and 2. 

\subsubsection{Attribute 7: Defense Capability Fidelity}
To address Threat 1, we can apply defense mechanisms at the WRKT element in the ground station component and the Pi5 element in the bus system component. At the WRKT element, we can: 
(i) employ traditional countermeasures against phishing and configure a firewall to filter traffic from the Internet to minimize the effects of phishing;
and (ii) segregate 
networks to separate command, telemetry, and payload data to limit the spread of a compromise. 
These mechanisms prevent an attacker from employing the attack vector leading to the attack consequence in the bus management function (Figure \ref{fig:bus_mngmnt_af}).

At the Pi5 element, we can: 
(i) establish a baseline of normal behavior for the solenoid valve and motor elements;
(ii) analyze propulsion and rotation 
commands 
to identify the ones that would produce anomalous behaviors;
(iii) prevent such commands from executing;
and (iv) send the log data to the ground station component for further analysis. These defense mechanisms can mitigate attacks
against the telemetry downlink function (Figure \ref{fig:telem_dwnlnk_af}).

To address Threat 2, we can apply defense mechanisms at the SDR elements in the ground and space segments. Specifically, we can:
(i) detect conspicuous jamming (e.g., overpowering of a single frequency);
(ii) identify inconspicuous (i.e., more sophisticated) jamming by inspecting the frequency, timing, power, angle of arrival, and polarity of signals for anomalies; and (iii) overcome jamming through a jamming-resistant communications scheme
(e.g., 
\cite{turner2024jamming}).
These defense mechanisms can mitigate attacks against 
the telemetry downlink and image capture functions (Figure \ref{fig:jamming_af}).

\begin{insight}
Analyzing mission-based attack fidelity (Attribute 6) could guide the hardening of missions. 
\end{insight}

\section{Future Research Directions}
\label{sec:discussion}

In this section, we discuss future research directions, including: enhancing attributes, enhancing the testbed, and designing experiments that can be conducted in the testbed.

\smallskip

\noindent{\bf Attributes Definition}. 
We defined 7 
attributes to characterize space cybersecurity testbeds. However, these 7 attributes may not be adequately comprehensive, meaning there may be other attributes that should be defined. A comprehensive set of attributes is important to comparing testbeds fairly. For instance, it is important to define attributes to characterize satellite constellations, such as the number of satellites in a constellation, the number of orbits involved, and the number of Inter-Satellite Links (ISL) accommodated by a testbed.
Moreover, future research needs to investigate how to quantify, for instance, threat model fidelity, which is nontrivial because there could be (infinitely) many threat models. We currently take a best-effort approach (i.e., counting the number of threat models that can be accommodated by a testbed).


\smallskip

\noindent{\bf Testbed Enhancement}. We plan to enhance the testbed in due course as this is a time-consuming process as open-source software is often buggy (e.g., cFS, Observation \ref{insight:buggy software}).
First, we currently have one fully functioning CubeSat, while we are building our second fully functioning CubeSat. We plan to build more CubeSats to formulate a constellation in the future. Second, our testbed currently lacks
a {\em battery management} element. 
The battery management 
element, which can be implemented in the OBC module, is important because it controls the energy distribution to the elements in a satellite; for instance, when the battery level is low, every element but the Pi5 
will be powered off until the battery level is above a threshold. The OBC module 
is a good candidate because it receives energy from the battery and redistributes the energy to the other modules in the satellite (see red arrows in Figure \ref{fig:system_model_element_level}). 
Third, we will investigate the cybersecurity implications of all the elements, modules, and components in our testbed. 
For instance, an attacker may use malware to manipulate the battery level to mislead the Pi5 element 
to power off the elements to cause the loss of communications.
Moreover, the workstation element may 
be compromised to maliciously maneuver a satellite to cause a 
conjunction.
Fourth, we will investigate cost-effective data collection because the same kind of data may be collected at multiple elements, but some elements may be more suitable than others. Fifth, we will aim to accommodate all the real-world space cyber attacks \cite{fritz2013satellite, us2019report,ear2023characterizing} in our enhanced testbed.

\smallskip

\noindent{\bf Designing Experiments}. The main purpose of building testbeds is to conduct experiments and validate/invalidate theoretic studies. We are investigating how to systematically design and conduct attack and defense experiments in our testbed. 
Moreover, we need to investigate how to leverage the testbed to conduct advanced analytics research. For instance, a set of desired properties for space cyber risk management tools are described in \cite{DBLP:journals/corr/abs-2402-02635}, which can guide the design and implementation of specific tools for space cyber risk management purposes. To validate the competency of such tools, we can leverage a testbed to measure the metrics that are associated with those properties, and then use these metrics to compare the competency of different tools. In particular, a testbed can be leveraged to measure metrics (e.g., \cite{Pendleton:2016,Cho16-milcom,XuSTRAM2018ACMCSUR,XuSciSec2021SARR}), while noting that the metrics defined for non-space cybersecurity environments (e.g., \cite{XuHotSoS2018Firewall,XuIEEETIFS2018-groundtruth,XuAgility2019}) may need to be adapted to the space cybersecurity environment. This will pave a way for validating advanced quantitative cybersecurity models \cite{
XuCybersecurityDynamicsHotSoS2014,XuBookChapterCD2019,XuMTD2020}.
For instance, the dynamics of malware that may be fighting against each other in satellite constellations may be different from its counterpart in enterprise networks
\cite{XuTDSC2012};
the dynamics of preventive and reactive cyber defense dynamics in satellite constellations may be different from its counterpart in enterprise networks \cite{XuTAAS2012,XuIEEETNSE2018,XuIEEEACMToN2019,XuTNSE2021-GlobalAttractivity};
the effectiveness of adaptive defense in satellite constellations may be different from its counterpart in enterprise networks \cite{XuTAAS2014}.

\section{Conclusion}
\label{sec:conclusion}

We have presented a framework for characterizing the fidelity of space cybersecurity testbeds. The framework defines attributes to characterize testbeds via the system models, threat models, and defenses that can be accommodated by them. We use the framework to guide us in designing and characterizing a concrete implementation of a testbed we have built. We described how to enhance our testbed, which could become a valuable asset to the community. 

\section*{Acknowledgments}
We thank Lynnane George and Nathaniel Means for their technical support in the process of building our satellites. This work is supported in part by the DoD UC2 program and NSF Grant \#2308142. The opinions expressed in the paper are that of the authors, and do not reflect the policy of any government agency in any sense.

\input{main.bbl}

\end{document}